\documentclass[twocolumn,amsmath,amssymb]{revtex4}

\usepackage{setspace}
\usepackage{graphicx}
\usepackage{dcolumn}
\usepackage{bm}
\usepackage{ae}

\usepackage{color}

\begin{document}

\newcommand{\eff}{{\text{eff}}}
\newcommand{\bare}{{\text{bare}}}
\newcommand{\back}{{\text{back}}}
\newcommand{\sat}{{\text{sat}}}
\newcommand{\res}{{\text{res}}}
\newcommand{\be}{\begin{equation}}
\newcommand{\ee}{\end{equation}}
\newcommand{\XXXX}{{\bf XXXX}}

\title{The renormalized jellium model for spherical and cylindrical colloids}

\author{Salete Pianegonda}%
\affiliation{Laboratoire de Physique Th\'eorique et Mod\`eles Statistiques 
(Unit\'e Mixte de Recherche UMR 8626 du CNRS), B\^atiment 100, Universit\'e de Paris-Sud, 91405 Orsay Cedex, France}
\affiliation{
Instituto de F\'{\i}sica, Universidade Federal do Rio Grande do Sul, CP 15051, 91501-970, Porto Alegre (RS), Brazil}

\author{Emmanuel Trizac}
\affiliation{CNRS; Universit\'e
Paris-Sud, UMR 8626, LPTMS, Orsay Cedex, F-91405}
\affiliation{Center for Theoretical Biological Physics, UC San Diego,                    
       9500 Gilman Drive MC 0374 - La Jolla, CA  92093-0374, USA}

\author{Yan Levin}%
\affiliation{
Instituto de F\'{\i}sica, Universidade Federal do Rio Grande do Sul, CP 15051, 91501-970, Porto Alegre (RS), Brazil}

\date{\today}

\begin{abstract}
Starting from a mean-field description for a dispersion of
highly charged 
spherical or (parallel)  rod-like colloids, we introduce the simplification of
a homogeneous background to include the contribution of other polyions to the 
static field created by a tagged polyion. The charge of this background is
self-consistently renormalized to coincide with the polyion effective charge,
the latter quantity thereby exhibiting a non-trivial density dependence, which directly enters into the equation of state 
through a simple analytical expression. 
The good agreement observed between the
pressure calculated using the renormalized jellium and Monte Carlo
simulations confirms the relevance of the 
{renormalized} jellium model for theoretical and experimental purposes and provides an
alternative to the Poisson-Boltzmann cell model since it is free of some
of the intrinsic limitations of this approach.

\end{abstract}

\maketitle

\section{Introduction}
\label{sec:intro}
In the realm of soft matter, a quantitative description of systems with a
non vanishing density of mesoscopic constituents (colloids) is a difficult
task whenever long range Coulomb interactions are present \cite{Belloni}. 
It is customary to 
introduce a suitably defined Wigner-Seitz-like cell to render the situation
tractable. This considerable simplification allows for the computation 
of various thermodynamic quantities (see e.g. \cite{Alfrey,Alexander}
and \cite{Hansenrevue,Deserno} for more recent accounts).
Transport properties may also be derived \cite{Kuwabara,Ohshima}.
Initially borrowed from solid state physics, the concept of a cell 
nevertheless appears fruitful to describe the phase behaviour 
of liquid phases (see e.g. \cite{LevinJPCM}). When a more microscopic
information such as effective interaction  is sought, 
there is however to date no evidence that the cell picture provides 
accurate answers, through the approximate procedures that have
been proposed to infer solvent and micro-ions averaged 
inter-colloid potentials of interaction \cite{Alexander}.
In the present paper, we adopt a 
more "liquid-state" viewpoint~\cite{LevinL} 
to describe the local and global screening properties of 
microscopic ions around
highly charged rod-like or spherical colloids, taking due account of the finite
density of colloids. Our approach bears a strong resemblance with 
a Jellium model put forward by Beresford-Smith, Chan and Mitchell \cite{Beresford},
with the important difference that the jellium under consideration 
here is  the  ``renormalized'' counterpart of that studied in \cite{Beresford}.
A preliminary account with emphasis on the sedimentation of charged
colloids has been published in \cite{Jellium} and we note that our approach
has been recently tested with some success on liposome dispersions
\cite{Haro,HaroJFCM06,CastanedaCM}.

The article is organized as follows. The model is defined in section 
\ref{sec:model} and illustrated in section \ref{sec:nosalt} where salt-free suspensions
are considered and the numerical procedure exemplified with charged spheres.
The cylindrical geometry is also addressed which allows to 
discuss the fate of classical Manning-Oosawa condensation 
phenomenon \cite{Manning,Manning2006} within the present
framework. The effects of an added
electrolyte are investigated in section \ref{sec:salt} and concluding remarks are 
presented in section \ref{sec:concl}. As will become clear below, our framework
provides a procedure to incorporate self-consistently charge
renormalization into the classical Donnan equilibrium description of suspensions.

\section{The model}
\label{sec:model}

When immersed in a polar solvent, mesoscopic ``particles'' release small
ions in the solution that, together with other micro-ions resulting from 
the dissociation of an added salt, form an in-homogeneous cloud
around each colloid.
Within the mean-field approximation to which we restrict here, neglect of
micro-ionic correlations allows to relate the local density of micro-species of
valency $z_i$ at point $\bm r$ to the local electrostatic potential $\varphi(\bm r)$
through $n_i({\bm r}) \propto \exp(-z_i e \beta \varphi(\bm r))$,
where $e$ is the elementary charge and
$\beta^{-1}=kT$ is the thermal energy \cite{Belloni,Levin}.
For any position of the $N$ colloids present, one needs to solve
the resulting Poisson-Boltzmann equation from which the electric potential
follows. This potential may then be inserted into the stress tensor \cite{Belloni}
to compute the force acting on the colloids.
Such a procedure, which
makes explicit use of the separation of time scales between colloids and
micro-ions, opens the way towards a complete description
of the statics and dynamics of the system, with e.g Monte Carlo or Molecular
Dynamics techniques to treat the colloidal degrees of freedom
\cite{Fushiki,Graz}. This treatment is however numerically 
particularly demanding and much insight is gained from further approximations
that map the original problem with $N$ colloids onto a one-colloid situation.
The cell model is an option, where the finite density of colloids is 
accounted for by an exclusion region. We propose here an alternative,
free of some of the limitations of the cell model,
that is equally simple to implement and solve.

A given colloid with bare charge $Z_\bare e$, assumed positive,
is tagged and fixed at the origin. The suspending medium
(solvent treated as a dielectric continuum with permittivity $\varepsilon$)
is taken as infinite, with a mean colloidal density $\rho$.
Following \cite{Beresford},
the colloids around the tagged particle are assumed to form 
an homogeneous background, with charge density 
$Z_\back e \rho$, so that the electrostatic potential around the tagged colloid
fulfills Poisson's equation
\be
\nabla^2 \varphi \,=\, -\frac{4 \pi}{\varepsilon} \, 
\left[Z_\back \, \rho \, e + \sum_i n_i^0 z_i e\, e^{-\beta e z_i \varphi}
\right]
\label{eq:Poisson}
\ee
where the summation runs over all micro-species and the concentrations
$n_i^0$ are determined either from electro-neutrality 
in the no salt case or from the osmotic equilibrium with a salt
reservoir in the semi-grand-canonical situation, both addressed
here.
At large distances ($r\to \infty$), the term in brackets
on the rhs of (\ref{eq:Poisson}) vanishes 
which imposes a  
value $\varphi_\infty$ for the potential far from the colloid,
that may be called a Donnan potential.
The key point in our approach is that unlike in \cite{Beresford},
$Z_\back\neq Z_\bare$:
the background  charge is not known {\it a priori} but is determined
self consistently as explained below. 

To illustrate the methodology, we consider a spherical colloid of radius $a$.
When $r \to \infty$, we may linearize  (\ref{eq:Poisson}) around
$\varphi_\infty$, which results in a Helmholtz equation indicating that 
\be
\varphi({\bm r}) \, \stackrel{r\to \infty}{\sim} \, \varphi_\infty  \,+\,
\frac{Z_{\eff} \, e}{\varepsilon(1+\kappa a)r} \, e^{-\kappa(r-a)}
\label{eq:ff}
\ee
where the characteristic distance $\kappa^{-1}$ is given by
\be
\kappa^2 \,=\, 4 \pi \sum_i  \frac{\beta e^2}{\varepsilon} n_i^0 z_i^2 \, e^{-\beta e z_i \varphi_\infty} = 4\pi \ell_B \sum_i z_i^2 n_i(\infty)
\ee
and $\ell_B=\beta e^2 /\varepsilon$ is the Bjerrum length.

For very low bare charges, the solution (\ref{eq:ff}) holds for all
distances with $Z_\eff=Z_\bare$, and one can consider that $Z_\back=Z_\bare$.
However, typical colloidal charges are such that 
$Z_\bare \ell_B/a \gg 1$, a regime for which 
counterions become strongly        
associated with the colloid and the charge renormalization effects
 \cite{Belloni98,PRL2002,JCP2002} can not be ignored.  
The counterion                                                 
condensation strongly affects the electrostatic far-field
so that the large distance signature involves an effective charge
[$Z_\eff$ in Eq. (\ref{eq:ff})] which significantly differs from the bare one.
As a result of non-linear screening, one has $Z_\eff \ll Z_\bare$
whenever $Z_\bare \ell_B/a \gg 1$.

At this point, the effective charge arising in (\ref{eq:ff}) is a function
of both the background and the bare charge, other parameters being fixed:
$Z_\eff=Z_\eff(Z_\back,Z_\bare)$. As far as a static description is
pursued, for sufficiently strongly charged colloids the         
bare charge is an irrelevant quantity far enough from the tagged
colloid, and we demand that $Z_\back$ coincides with $Z_\eff$, which 
best characterizes the background charge resulting from smearing
out the other colloids contribution. We therefore
enforce the self-consistency constraint 
\be
Z_\back \,=\, Z_\eff(Z_\back,Z_\bare)
\ee
to compute the {\it a priori} unknown background charge.
As we shall see below, this condition is readily implemented numerically
and for a given $Z_\bare$, leads to a unique value for 
$Z_\back=Z_\eff$. This value is density dependent, which is also
the case of the inverse screening length $\kappa$.
Indeed, $\varphi_\infty$ depends on $\rho$ \cite{rque}
through the electro-neutrality
condition
\be
Z_\back \, \rho \, e + \sum_i n_i^0 z_i e\, \exp\left(-\beta e z_i \varphi_\infty\right)=0
\label{eq:electroneutrality}
\ee
which translates into a $\rho$ dependence for $\kappa$. 
Considering now two colloids in the weak overlap
approximation (i.e. not too close), the effective potential of interaction
will take a DLVO form \cite{Belloni,Levin} with effective
parameters $\kappa$ and $Z_\eff$. 

The procedure outlined here incorporates non linear screening
together with finite $\rho$ effects. It is best suited to describe 
low density systems since the 
colloid-colloid pair distribution function $g_{cc}$
is implicitly considered to be unity for all distances. 
This reduction,  which has non trivial consequences,
is certainly of little relevance for high density suspensions
for which the cell model is presumably a better approximation.

Before illustrating the method, we briefly consider the pressure
in the system, that is given by the densities of micro-ions
far from the tagged colloid:
\be
\beta P \,=\,  \sum_i n_i(\infty) = \sum_i n_i^0 \, \exp\left(-\beta e z_i \varphi_{\infty} \right).
\label{eq:P}
\ee
The colloidal contribution is explicitly discarded \cite{Jellium}. This is well justified
in the low salt limit, which is a regime of counterions dominance
provided that $Z_\bare \gg 1$, which is easily achieved in practice.

\section{The no salt limit}
\label{sec:nosalt}
 
\subsection{Spherical colloids}
The simplest situation to investigate is that of de-ionized suspensions (no salt).
For simplicity, we consider counterions as monovalent. From (\ref{eq:Poisson})
it follows that the dimensionless potential $\phi = \beta e \varphi$ obeys the equation
\be
\frac{d^2 \phi}{d \tilde r^2} \,+\, \frac{2}{\tilde r} \frac{d \phi}{d\tilde r} \,=\,
3 \eta \, \frac{Z_\back \ell_B}{a} \, \left(e^{\phi} -1 \right)
\label{eq:sphnosalt}
\ee
where $a$ is again the radius of the tagged particle from which the
dimensionless distance $\tilde r=r/a$ is defined, and $\eta=4 \pi \rho a^3/3$ is the
volume fraction. The boundary conditions are 
\begin{eqnarray}
\phi \to 0 &&\quad \hbox{for} \quad \tilde r \to \infty 
\label{eq:bc1}\\
 \frac{d \phi}{d\tilde r} = -\frac{Z_\bare \ell_B}{a} &&\quad \hbox{for} \quad \tilde r =1.
 \label{eq:bc2}
\end{eqnarray}
In writing (\ref{eq:sphnosalt}), use has been made of the 
(global) electro-neutrality constraint
$n_-^0 e \exp(\phi_\infty) = Z_\back \rho e$ with the choice $\phi_\infty=0$. 
For all values of $Z_\back$,  the far-field of $\phi$ is governed by
$\kappa$ such that
\be
(\kappa a)^2 \,=\, 3 \eta \frac{Z_\back \ell_B}{a}.
\label{eq:kappasphnosalt}
\ee

The above system is solved following similar lines as in \cite{Langmuir2003}.
We summarize here the main steps.
In practice, equation (\ref{eq:sphnosalt}) is solved numerically for
a finite system $\tilde r \in [1,\widetilde R]$, where $\tilde R$ needs to be large enough
(that is $\kappa a \widetilde R \gg 1$ but note that $\kappa$ is not known initially
but follows once the background charge is known). 
a) The first and important step is to 
rephrase the boundary value problem at hand as an initial value problem 
with boundary conditions $\phi'(\widetilde R)=0$ (to ensure electro-neutrality)
and varying $\phi(\widetilde R)$. The volume fraction $\eta$ is fixed and
the background charge $Z_\back$  
first assigned a guess value, to be modified iteratively (see below).
If $\phi(\widetilde R)$ is small enough,
the system then admits a solution. 
b) From this solution, one computes $\phi'(\tilde r=1)$ 
to know the corresponding bare charge. 
c) Changing $\phi(\widetilde R)$  \cite{rque2}, the targeted value
$\phi'(\tilde r=1)=-Z_\bare \ell_B/a$ is readily found  by iteration. 
d) The screening quantity $\kappa$ 
is subsequently computed from (\ref{eq:kappasphnosalt}) and the
effective charge associated to the particular couple $(Z_\back,Z_\bare)$
is deduced from  the large $\tilde r$ behavior of $\phi$ [e.g
one needs to observe a clear-cut plateau for 
$[\phi(\tilde r)-\phi(\widetilde R)] e^{\kappa a \tilde r} \tilde r $ plotted as a function of $\tilde r$
in the range $1 \ll \tilde r <\widetilde R$]. The first iteration ends here,
and the procedure is repeated with the $Z_\eff$ obtained as the next
trial value for $Z_\back$. 
Alternatively,  one may sample several trial values for $Z_\back$
and plot $Z_\eff$ versus $Z_\back$. As may be observed in Fig.
\ref{fig:Zsph1} where such a plot is displayed, 
the dependence of $Z_\eff$ on $Z_\bare$
is very mild, which means that convergence towards 
$Z_\back=Z_\eff$ is achieved in a few steps.
In the (artificial) limit where $Z_\back \to 0$, the problem at hand
reduces to an unscreened one (governed by Laplace equation)
with solution $\phi = Z_\bare \ell_B/r$: there is therefore no
renormalization of effective charge so that $Z_\eff \to Z_\bare$
(see Fig. \ref{fig:Zsph1}). {The inset shows how the 
self-consistent background charge is determined, the other points
being unphysical.}

{In the limit of a diverging bare charge, the procedure is well
behaved and yields a finite self-consistent effective charge.  From the
previous discussion, we expect $Z_\eff$ to diverge at small $Z_\back$, which
is indeed the case (not shown).}

{Once the physical solution to the problem has been located
(inset of Fig. \ref{fig:Zsph1}), 
various quantities such as the pressure
may be computed. In the remainder, we will use the terms 
``effective'' and ``background'' charges to refer to the self-consistent
solution as obtained in Fig. \ref{fig:Zsph1}:
 $Z_\eff=Z_\back$ is therefore
a function of $Z_\bare$ and volume fraction (possibly also salt concentration,
see section \ref{sec:salt}). For a particular density, this function 
is shown in Fig. \ref{fig:Zsph3}. After the initial linear regime,
where no renormalization takes place, the effective charge slowly reaches
a saturation plateau as $Z_\bare \to \infty$. For this specific density
($\eta=10^{-2}$) the effective charge saturates at $Z_\eff \ell_B/a \simeq 6.6$.}
The saturation phenomenon observed here is strongly reminiscent of
that observed in the classical Poisson-Boltzmann approach
(either in a cell, or in an infinite medium \cite{Alexander,JCP2002,Tellezsat}).
To assess quantitatively the possible difference with the results
obtained within the cell model, we compare in Fig. 
\ref{fig:sphcompcellns} $Z_\eff$ derived in the cell 
\cite{Alexander,Langmuir2003} to its {renormalized} jellium counterpart.
Both charges differ by a notable amount for $\eta >10^{-3}$
while the agreement at very low density is meaningless, 
and follows from a divergence of the saturation effective charge
in both models: non-linear effects disappear when 
$\eta \to 0$, 
so that $Z_\eff \to Z_\bare$. This is a peculiarity of systems with colloidal spheres and counterions only, and it turns out that the behavior of charged cylinders is quite 
different, see below.

Under the de-ionized conditions studied here, the pressure
takes the simple form $\beta P = Z_\eff \rho$, whereas 
the corresponding expression in the cell model is less explicit
and does not directly involve the effective charge.
Remarkably, although there is a significant difference
between the effective charges calculated within {the Poisson-Boltzmann cell
model (PBC) and the renormalized jellium},
the pressures calculated using the two models are identical for $\eta<0.1$,
Fig. \ref{fig:Zsphpressns}. A similar agreement is found
at saturation \cite{Jellium}. We add here that a comparison 
between the {renormalized} jellium equation of state and the ``exact'' results of the primitive
model obtained using the Monte Carlo
computations has been reported in \cite{Jellium},
with excellent agreement (the corresponding density range
is quite low such that PBC and {renormalized} jellium predictions agree). 

Before concluding this section, we emphasize that one must 
carefully check that the results obtained do not depend on the particular
value chosen for the cutoff $\widetilde R$, e.g. by repeating 
the analysis with an increased cutoff.

 \subsection{Cylindrical colloids}
 
 Consider a nematic phase of parallel infinite rods {($L \to \infty$)} with
 line charge $\lambda_\bare$ (therefore no
 positional order in the plane perpendicular to the main axis).
 We may repeat the previous approach, tagging a given rod of radius $a$
 and model the effects of the other rods  by an homogeneous
 background, with line charge $\lambda_\back$.
 While the definition of $\kappa$ is unaffected compared to the
 spherical case, the far-field potential now takes the form:
 \be
 \phi(\tilde{r}) \,=\, \phi_{\infty} \,+\, 2 \lambda_{\eff} \ell_{B} \,
 \frac{K_{0}(\kappa a \tilde{r})}{\kappa a K_{1}(\kappa a)},
 \ee
 where $K_0$ (resp. $K_1$) denotes the zeroth (resp. first)  
 order modified Bessel function of the second kind.
In the spirit of the consistency requirement of section \ref{sec:model},
we impose $\lambda_\back=\lambda_\eff$ where again,
$\lambda_\eff$ follows from the large distance behavior of
the solution to Poisson's equation with background charge 
$\lambda_\back$:
\be
\frac{d^2 \phi}{d \tilde r^2} \,+\, \frac{1}{\tilde r} \frac{d \phi}{d\tilde r} \,=\,
4 \eta \, \lambda_\back \ell_B\, \left(e^{\phi} -1 \right).
\label{eq:cylnosalt}
\ee
Here the volume fraction is $\eta = \pi a^2\, n_{_{\text{2D}}}$ where 
$n_{_{\text{2D}}}$ is the mean surface density of rods (in the plane perpendicular
to their axis). The boundary conditions are the same as (\ref{eq:bc1}) and (\ref{eq:bc2}),
and the numerical method identical to that used in the spherical case.

{The effective charges calculated using the cell and the renormalized
jellium models are compared in Fig. \ref{fig:lambanosat} for
$\lambda_{bare}\ell_B=1$. The inset corresponds to the
saturation regime where $\lambda_\bare$ is very large ($\lambda_\bare \to \infty$). We observe a 
substantial disagreement between the two effective charges.}
On the other hand, in the small bare charge regime where
non-linear effects are not at work, both quantities coincide (not shown),
which is a signature of 
whenever non-linear effects come into
play (i.e. outside the small bare charge linear regime, which is the
case for both figures).
Beyond these differences, Manning-Oosawa 
condensation \cite{Manning,Manning2006}, which is a key feature of 2D
electrostatics, is shared by both PBC and {renormalized} jellium models. As the colloid
density is decreased ($\eta \to 0^+$), the effective charge becomes independent
of the bare one, provided $\lambda_\bare$ exceeds the critical threshold
$1/\ell_B$. This feature is illustrated in Fig. \ref{fig:Manning}.
At the saturation plateau and again for $\eta \to 0^+$, 
one has $\lambda_\eff \ell_B\simeq 0.47$,
a value that will be refined below.

To be more quantitative, it is furthermore natural to compare
the corresponding functional forms of effective charges
versus bare ones, and versus density in both PBC (where it can be computed
analytically) and {renormalized} jellium models, where this information is accessed numerically. 
To this end, we reconsider the 
analytical results obtained in \cite{JCP2002, Langmuir2003}
where the effective charge in the cell model following Alexander {\it et al.}
prescription \cite{Alexander} reads:
\begin{eqnarray}
\lambda_{\eff} \ell_{B}&=&\frac{1}{2}K_{PB}^{2}aR_{ws}\{I_{1}(K_{PB}R_{ws})K_{1}(K_{PB}a)\nonumber \\
& & {}  -I_{1}(K_{PB}a)K_{1}(K_{PB}R_{ws})\},
\label{eq:PBCMann}
\end{eqnarray}
with standard notation for the Bessel functions.
Here $R_{ws}\equiv a \eta^{-1/2}$ is the radius of the cell and $K_{PB}$
is the inverse screening length related to the micro-ionic density at the cell 
boundary \cite{JCP2002}, which can be computed explicitly 
from the analytical solution of \cite{Alfrey}. After some algebra, we find,
to leading order in density that when $\lambda_\bare > 1/\ell_B$
\begin{equation}
\lambda_{\eff}^\sat \ell_{B} \,\stackrel{\eta \to 0^+}{\sim} \, \frac{\sqrt{2}}{2}I_{1}(\sqrt{2}) \, +\, \pi^2
\frac{I_{0}(\sqrt{2})+\sqrt{2}I_{1}(\sqrt{2})+I_{2}(\sqrt{2})}{(2\xi-\log(\eta))^2},
\label{eq:form}
\end{equation}
where $\xi = \lambda_\bare/(\lambda_\bare -1/\ell_B )$. 
We note that the leading term $ \sqrt{2}I_{1}(\sqrt{2})/2 \simeq 0.63$
differs from the value found in the {renormalized} jellium ($\simeq 0.47$, see Fig. 
\ref{fig:Manning}). 
Moreover, Eq. (\ref{eq:form}) also 
suggests a fitting form to describe the saturation plateau in the low density 
regime of the {renormalized} jellium model:
\be
\lambda_\eff^\sat \ell_B \,\stackrel{\eta \to 0^+}{\sim} \, A + \frac{B}{(C-\log \eta)^2}. 
\ee
{The values of $A$, $B$ and $C$ can be obtained using a numerical fit. We find
that in the saturation limit $A\simeq 0.471$, $B\simeq16.87$ and $C\simeq
0.843$ give an excellent agreement with the numerical data. We have also checked that 
an equally good agreement is found at lower bare charges,
such as $\lambda_\bare \ell_B=4$, but with different values of $A$, $B$ and $C$.}
We conclude here that both models are described by the same
limiting law for low densities, at least beyond the condensation threshold.

It is of interest to resolve the condensate structure once the counterion
condensation has set in. A useful measure of the condensate thickness
is provided by the so-called Manning radius $R_M$ \cite{Gueron}
that has been recently
worked out in the infinite dilution limit and for low salt content 
\cite{OS,Manning2006}: in practice, the integrated charge 
per unit length
$q(r)$ around
a rod has an inflection point
at $r=R_M$, when plotted as a function of $\log r$. This is exactly
the point where $q(R_M)\ell_B/e=1$.
We expect a similar
behavior for {the renormalized} jellium, given that in the vicinity of highly charged rods, 
the (largely dominant) counterion
distribution should not be sensitive to the difference between
a uniform background as in the {renormalized} jellium model, and coions
as in the situation worked out in \cite{Manning2006}. The lower inset of
Figure \ref{fig:man1} shows that this is indeed the case.
In addition, from the analytical expressions derived
in \cite{Manning2006} and the fact that the relevant screening parameter
reads here $(\kappa a)^2 = 4 \eta \lambda_\eff \ell_B$, we expect
the scaling $\kappa R_M \propto (\kappa a)^{1/2}$, more precisely
\be
R_M \, \stackrel{\eta \to 0^+}{\propto} \, a\, \eta^{-1/4} 
\exp\left(-\frac{1}{2(\lambda_\bare \ell_B-1)}
\right).
\label{eq:mann}
\ee
The dependence of $R_M$ on both density and bare charge embodied
in Eq. (\ref{eq:mann}) is fully supported by the numerical data, see
Fig. \ref{fig:man1}.

Finally, and much like for spherical colloids, 
there is a good agreement between the osmotic pressure calculated using the cell model
and the renormalized jellium approximation, in spite of the different effective charges, 
 see Fig. \ref{press}. Discrepancies are observed only for volume fractions
$\eta>0.06$ and the agreement seems to be better at high charges.

\section{Effects of added salt}
\label{sec:salt}

In this section, we consider systems dialyzed against an electrolyte reservoir
with the monovalent salt concentration $c_s$. The corresponding screening
parameter is 
$\kappa_{\res}^2=8\pi\ell_{B}c_{s}$. 
It is convenient to choose the reference potential so that micro-ionic
densities are $n_\pm({\bm r}) = c_s \exp[\mp \phi({\bm r})]$,
where the counterions are assumed to be monovalent.
Using Eq. (\ref{eq:electroneutrality}), the potential at infinity becomes  
\be
\phi_\infty \,=\, \hbox{arcsinh}\left(\frac{Z_\back \rho}{2 c_s} \right).
\ee

It is important to keep in mind that  
$n_{\pm}({\bf r})$ are not the physical microion 
densities, but are the effective (renormalized) quantities satisfing 
\be
 \int d {\bm r} [n_+({\bm r}) -n_-({\bm r}) +Z_\eff
\rho]= - Z_\bare.
\ee
{Since the renormalization does not affect coions, their concentration inside the jellium with
one colloid  fixed at $\bm r=0$ is}
\be
C_+ = \frac{1}{V}\int  d {\bm r} n_+({\bm r}),
\ee
where it is understood that $V$ denotes the measure of a large 
volume centered at $\bm r=0$.
The  concentration of counterions, $C_-$, then
follows from the
overall charge neutrality inside suspension, 
$C_- = C_+ +Z_\bare \rho $.

Far from colloid, $n_+({\bm r})$ saturates at the bulk value 
$\widetilde n_+$, so that in the 
thermodynamic limit ($V \rightarrow\infty$)
\be
C_+ =\widetilde n_{+}.
\label{eq1}
\ee
Similarly, for $V\to \infty$
\be
\frac{1}{V}\int d {\bm r} [n_+({\bm r}) -n_-({\bm r}) +Z_\eff
\rho]= \frac{- Z_\bare}{V}\rightarrow 0 \;,
\ee
which means that
\be
 C_+- \frac{1}{V}\int d {\bm r}n_-({\bm r})+Z_\eff\rho=0 \;.
\label{eq3}
\ee
The charge neutrality  allows us to 
rewrite Eq.~(\ref{eq3}) as
\be
 \frac{1}{V} \int  d {\bm r} n_-({\bm r})=C_--(Z_\bare-Z_\eff)\rho \;.
\label{counter}
\ee
Eq.~(\ref{counter}) provides a suggestive 
interpretation of $n_-({\bm r})$ 
as the local density of free (uncondensed) counterions.  
Far from colloid, $n_-({\bm r})$ saturates at its bulk value 
$\widetilde n_-$, and in 
thermodynamic limit Eq.~(\ref{counter}) reduces to
\be
C_-= \widetilde n_-+(Z_\bare-Z_\eff)\rho.
\label{eq2}
\ee 
Eqs.~(\ref{eq1}) and (\ref{eq2}) allow us to calculate the ionic content
inside a suspension  dialized against a salt reservoir.  This
is particularly useful when comparing  the results of the
{renormalized} jellium model, which is grand canonical in electrolyte, 
with the Monte Carlo simulations, which are usually performed in a
canonical ensemble.
Knowledge of the assymptotic potential allows us to obtain the  
concentrations of coions and {\it free} 
counterions inside the suspension,
\be
\widetilde n_{\pm} \,=\, c_s \exp(\mp \phi_\infty).
\ee
These  are precisely the densities that
govern screening within {the renormalized} jellium, 
$\kappa^2 = 4 \pi \ell_B (\widetilde n_++\widetilde n_-)$.

\subsection{Spherical colloids}

In spherical geometry, Poisson equation (\ref{eq:Poisson}) now 
takes the form
\be
\frac{d^2 \phi}{d \tilde r^2} \,+\, \frac{2}{\tilde r} \frac{d \phi}{d\tilde r} \,=\,
(\kappa_\res a)^2 \sinh \phi -
3 \eta \, \frac{Z_\back \ell_B}{a} .
\label{eq:PSph}
\ee
We again solve it numerically as a boundary value problem in a (large enough) 
finite cell with vanishing $\phi'$ at the boundary, increasing gradually
the boundary potential from the value
\be
\phi_\infty \,=\,
\hbox{arcsinh}\left[\frac{ 3 \eta \, Z_\back  \ell_B/a}{(\kappa_\res a)^2} 
\right],
\ee
which corresponds to a vanishing bare charge.

Linearizing Eq. (\ref{eq:PSph}) around $\phi_\infty$, it can be seen that at
large distances the potential takes the form of Eq. (\ref{eq:ff}), with a 
screening constant $\kappa$ given by
\be
(\kappa a)^4 \,=\, (\kappa_\res a)^4 + \left(
\frac{3 \eta Z_\back \ell_B}{a}     \right)^2.
\ee
For highly charged colloids and typical salt conditions, 
the corresponding density dependence is shown
in Fig. \ref{fig:kapsphsalt}, while the effective charge (deduced from the condition
$Z_\eff = Z_\back$) is displayed in Fig. \ref{fig:Zsphsalt}. 
When $\eta \to 0$, both quantities coincide with the infinite dilution limit
of the traditional
PB theory, as they should. The increase of $\kappa$ 
with the density of colloids reflects the increasing importance of counterion
screening. The effective charge shows a non-monotonous behaviour with respect to
density.

To compute the osmotic pressure, we subtract the reservoir pressure
($2 c_s kT$) from the expression (\ref{eq:P}). Moreover, 
it should be remembered that such a relation only provides the
ionic contribution to the pressure. In the presence of salt and at low
colloidal density
this contribution becomes smaller than the colloidal one.
The vanishing of the microion contribution to pressure is exponential in the cell model, while it is 
algebraic for the jellium. Both models should then strongly disagree in the
low density limit.
To mimic the colloidal contribution, we add the ideal gas term $\rho k T$ to
(\ref{eq:P}), so that 
 the resulting osmotic pressure reads
\begin{equation}
\beta \Pi =\rho+\sqrt{Z_{\eff}^2\rho^2+4c_{s}^2}-2c_{s}.
\label{pressalt2}
\end{equation}
In the no salt case, addition of the ideal term is irrelevant since it is
always much smaller than
the micro-ionic one, provided that $Z_\eff$ is large
enough (this is the case for highly or even weakly charged colloids
since $a \gg \ell_B)$.   Moreover,
addition of the ideal gas term breaks the scaling form valid in the
no salt case where $a^2 \ell_B \beta P $ only depends on 
$\eta$ and reduced charge $Z_\bare \ell_B/a$. We therefore show the
osmotic pressure in Fig. \ref{fig:pressphsel} for two values
of colloid radius, within both the PBC and the {renormalized} jellium 
frameworks. Apart from  the expected deviations at small densities,
one observes compatible values at higher $\eta$.

There exist relatively little simulational data for the primitive model with salt, 
where the bare Coulomb interactions between all charged species
--colloids and micro-ions--
are taken into account (with still an implicit solvent). A reference 
equation of state  with salt is provided in
\cite{Lobaskin},
with the simplification  of a Wigner-Seitz cell, but explicit micro-ions.
The simulations were performed in canonical ensemble with
fixed salt content. 
The amount of added salt is characterized by a ratio
of the overall added cation charge to the overall macroion charge,
$\beta_{L}=C_{+}/(Z_{\bare} \,\rho).$
We compute the densities $C_{\pm}$ corresponding to a given salt
content as discussed in section \ref{sec:salt}. In Fig. \ref{fig:Lobaskin}
the osmotic pressure $\beta P/\rho_{t}$ is plotted as a function of $\beta_{L}$, where $\rho_{t}$
is the total density of ionic species. As in the case of salt-free
suspensions the pressures calculated using the PBC and the {renormalized} jellium are in good agreement.

\subsection{Rod-like colloids}

For completeness, we briefly report here results for cylindrical geometry.
Unlike salt-free case where $\lambda_\eff$ is a monotonic function 
of density, a minimum appears in the {renormalized} jellium curve shown in Fig. \ref{fig:k0k1}. 
The agreement between PBC and {renormalized} jellium at low $\eta$ signals
the region where the system is salt dominated (the colloid density 
is too low and, consequently, counterions do not participate in screening).
Conversely, the inset indicates the density range where counterions do 
dominate :  for $\eta > 10^{-1}$, the results become independent of the reservoir 
ionic strength and coincide with those obtained in the no salt limit.

{Finally, the pressure (\ref{pressalt2}) for cylindrical colloids is given by
\begin{equation}
4\pi\ell_B a^2\beta \Pi =4\eta\frac{\ell_B}{L}+\sqrt{(4\lambda_{eff}\ell_B\eta)^2+(\kappa_{res}a)^4}-(\kappa_{res}a)^2.
\label{pressaltsgi}
\end{equation}
Note that for infinite polyions ($L\to \infty$), the first term on the right
hand side of Eq. (\ref{pressaltsgi}) vanishes. In Fig. \ref{presscylsel2} we
plot the equation of state for polyions of $\lambda_\bare\ell_B=2$. One should note a strong
disagreement between the equation of state obtained using the renormalized
jellium model and the PBC theory. In the case of cylindrical polyions  the
disagreement is exacerbated by the fact that the ideal gas contribution to
the equation of state, Eq. (\ref{pressaltsgi}), vanishes in the limit of $L\to
\infty$ considered in this work. For small $\eta$, the behavior predicted by
the renormalized jellium model is more realistic than that of the PBC.}

\section{Conclusion}
\label{sec:concl}


Starting from a mean-field description in which a dispersion of $N$
spherical or rod-like polyions is treated using a $N$-body
Poisson-Boltzmann theory \cite{rque3}, we have introduced the simplification of
a homogeneous background to include the contribution of other colloids to the 
static field created by a tagged colloid. The charge of this background is
consistently renormalized to coincide with the effective charge 
governing the far-field potential. This results in a non-trivial density dependence
of the effective colloidal charge, which directly enters into the equation of state 
through a simple analytical expression. The good agreement observed between the
pressure calculated using the renormalized jellium and the Monte Carlo
simulations confirms the relevance of the {renormalized}
jellium model for theoretical and experimental purposes and provides an
alternative to the Poisson-Boltzmann cell approach.
 Furthermore, we note that the effective
charge calculated using the renormalized jellium model should be more relevant
for the study of the effective interaction between the colloids than its 
Poisson-Boltzmann cell (PBC)
counterpart. This is particularly the case since 
at finite colloidal density, the DLVO potential arises naturaly within the
jellium formalism, while it has to be introduced extraneously within the PBC.
In this work, we have left 
untouched the question whether the pair potential calculated using jellium
is a potential of mean-force
or an effective pair potential (following the terminology of \cite{Belloni}). 
Further work is required to answer this question.

In a cylindrical geometry, the present approach implicitly subsumes an
alignment between infinite rods --which is also a prerequisite for the analysis of \cite{Alfrey}--
but contrary to the crystalline structure underlying the introduction of 
the cell model, we consider here systems with no positional order
for the rods. As for spherical colloids the pressure in both approaches 
is in good agreement up to relatively high densities, whereas the
effective charges differ significantly. We have also shown that the scenario for counterion 
condensation is similar to that of the cell picture.

Our approach, which is best suited to describe systems with low macro-ions 
densities, may be easily extended to the case of asymmetric electrolytes.
One interesting aspect of {the renormalized} jellium is that the description of colloidal mixtures 
(macro-ions with different sizes and charges) appears to be as straightforward
as for the mono-disperse systems reported here.
This is an important difference with the cell approach, which cannot be
easily extended to such systems.

Among the possible refinements, it is possible to 
consider an in-homogeneous jellium with, again, renormalized charge.
This should allow to extend the relevant range of densities where
the model holds. Another interesting extension deals with
the derivation of electro-kinetic properties. Work along these lines
is in progress.

\section{Acknowledgments}  The financial support of Capes/Cofecub is gratefully acknowledged. 
We would also like to thank M. Deserno, J. Dobnikar, H.H. von Gr\"unberg,
R. Casta\~neda Priego and L. Belloni for useful discussions.

\newpage



\newpage
\section{Captions}

\begin{figure}[h]
\centering
\resizebox{9cm}{!}{\includegraphics[clip]{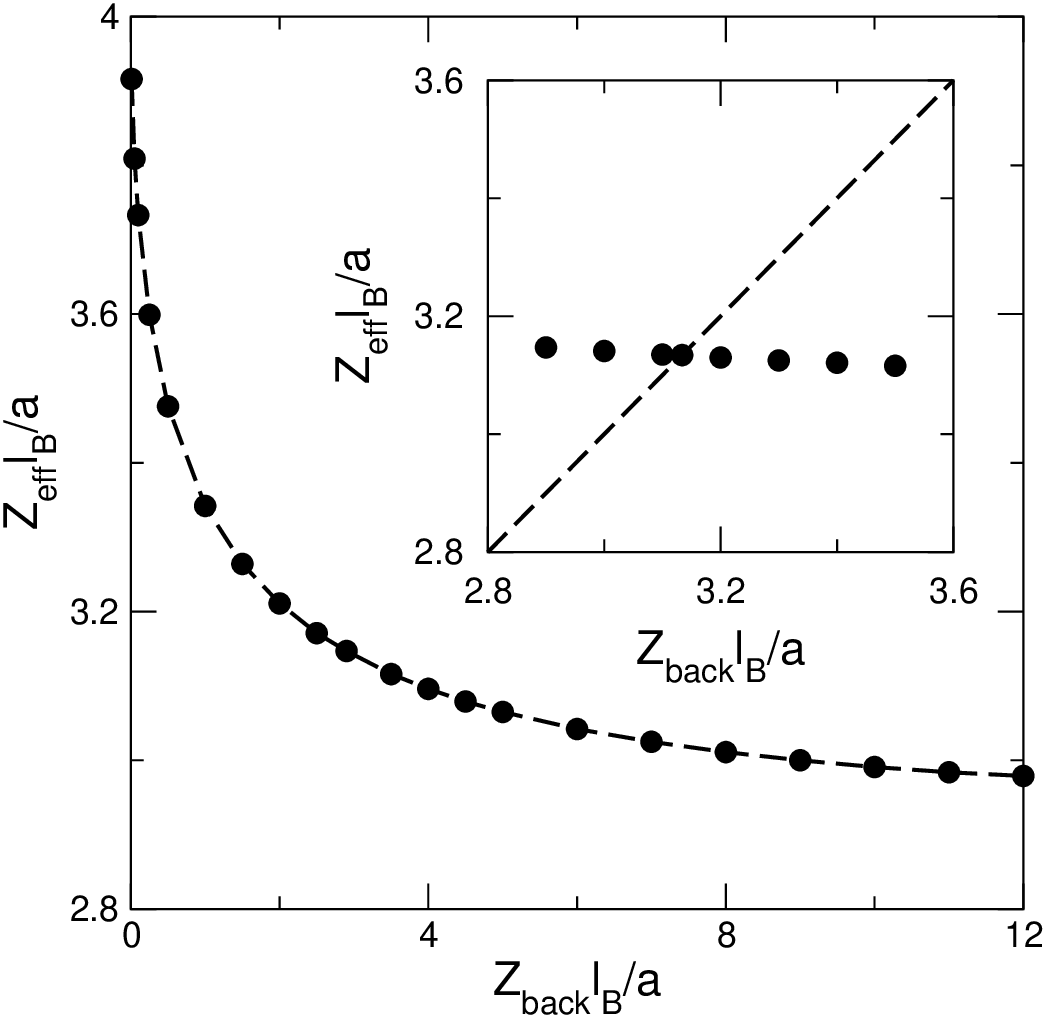}}
\caption{}
\label{fig:Zsph1}
\end{figure}

Fig. (\ref{fig:Zsph1}): The effective charge as a function of 
background charge for $\eta=10^{-2}$ and $Z_{\bare}\ell_{B}/a=4$ 
(spherical colloids, no salt). The physical 
solution $Z_\eff=Z_\back$
to the problem is the point of intersection with the first
bisectrix (see inset, where a magnification of the relevant part
of the main graph is shown).\\

\begin{figure}[h]
\centering
\resizebox{9cm}{!}{\includegraphics[clip]{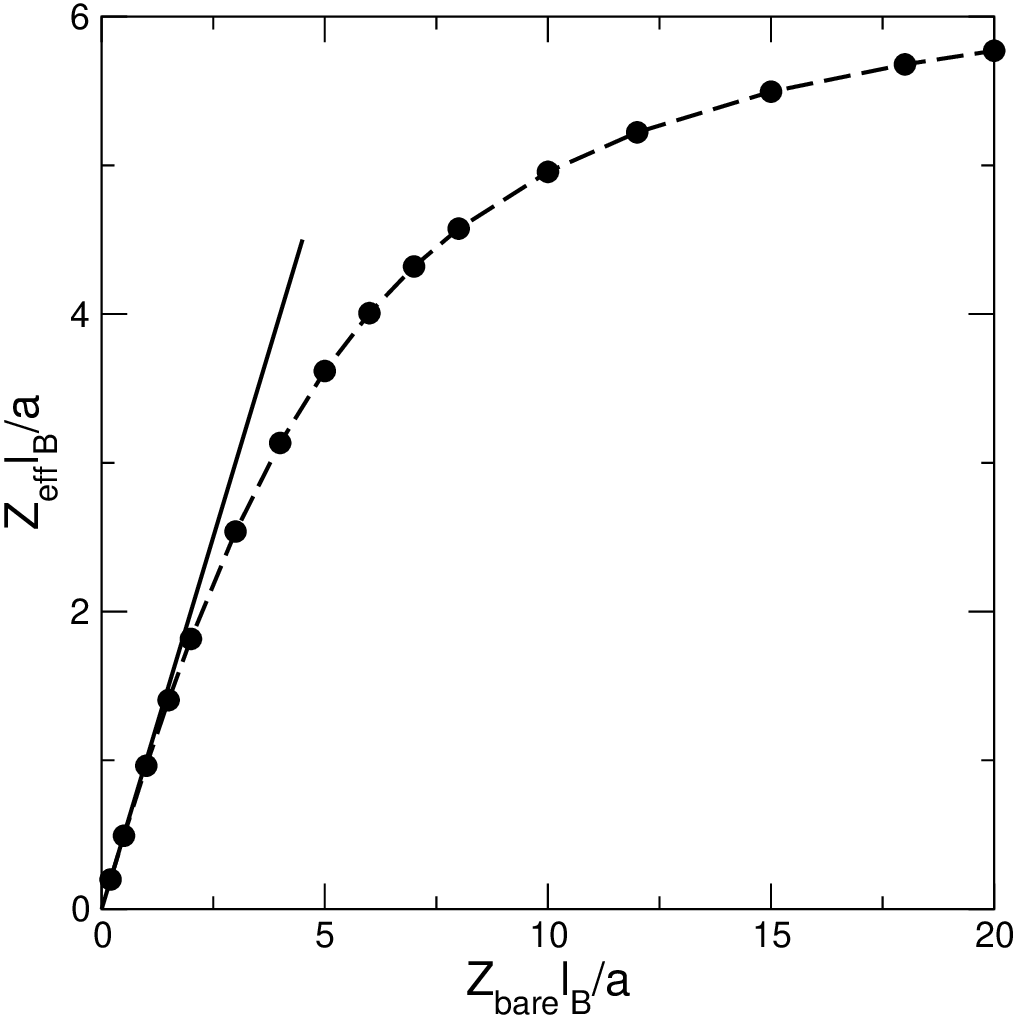}}
\caption{}
\label{fig:Zsph3}
\end{figure}

Fig. (\ref{fig:Zsph3}): The effective charge (or equivalently, the background charge)
as a function of 
bare charge for $\eta=10^{-2}$ (spherical colloids, no salt).
The line has slope unity to emphasize the initial
``Debye-H\"uckel'' regime.\\

\begin{figure}[h]
\centering
\resizebox{9cm}{!}{\includegraphics[clip]{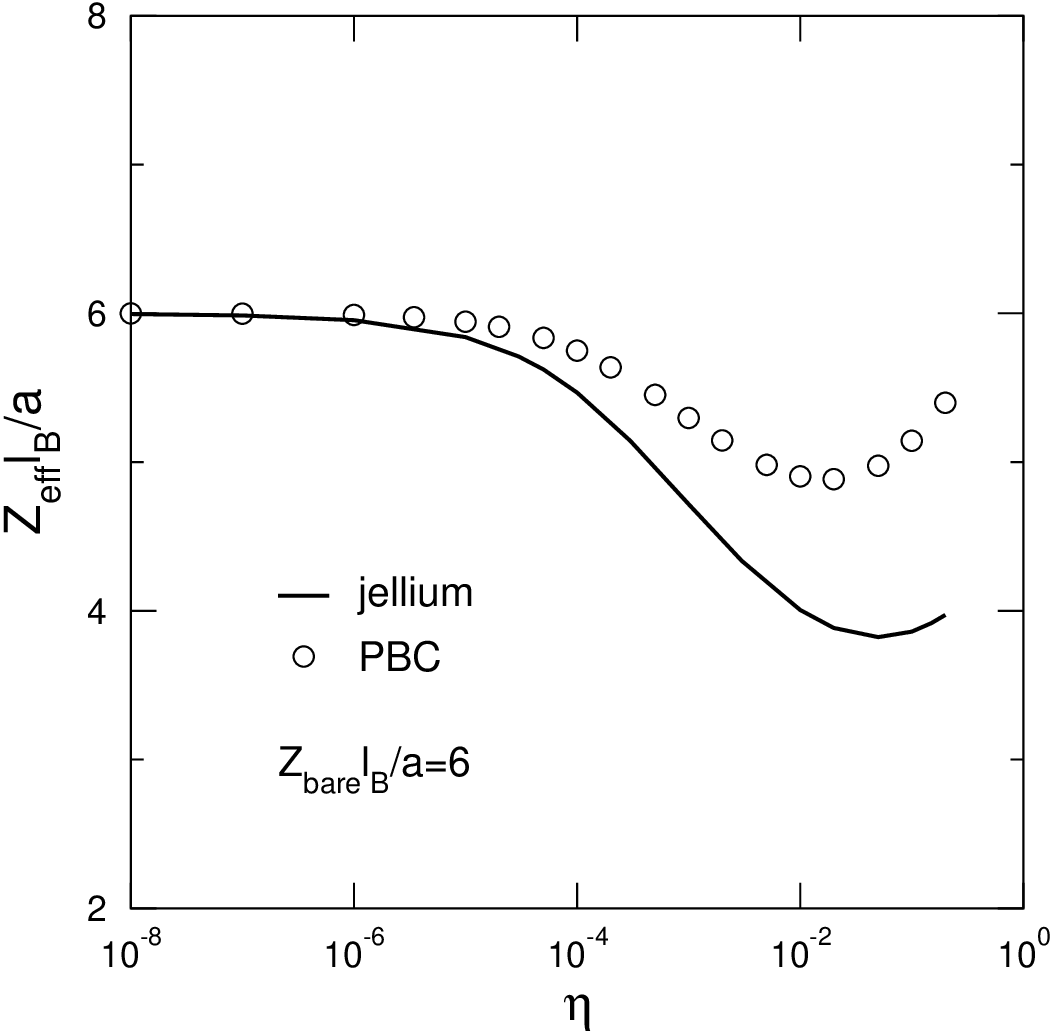}}
\caption{}
\label{fig:sphcompcellns}
\end{figure}

Fig. (\ref{fig:sphcompcellns}): Comparison between the effective charges 
within the Poisson-Boltzmann cell (PBC) and the renormalized jellium model,
as a function of the volume fraction. Here $Z_{\bare} \ell_{B}/a=6$ 
(spherical colloids, no salt).\\

\begin{figure}[h]
\centering
\resizebox{9cm}{!}{\includegraphics[clip]{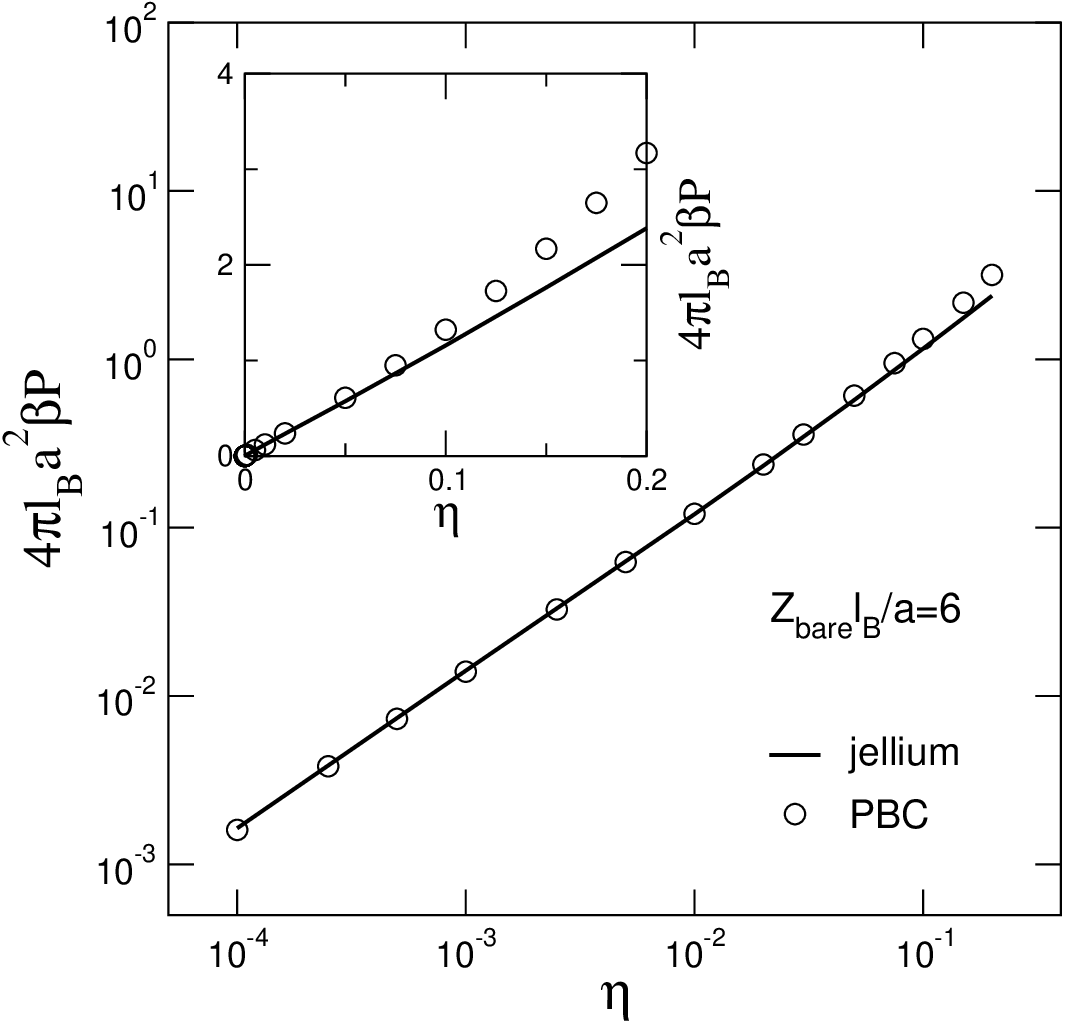}}
\caption{}
\label{fig:Zsphpressns}
\end{figure}

Fig. (\ref{fig:Zsphpressns}): Pressure as a function of volume fraction within
the cell and the renormalized jellium model,
on a log-log scale. Here,  $Z_{\bare}l_{B}/a=6$ (spherical colloids, no salt). 
The inset shows the same data on a linear scale.\\

\begin{figure}[h]
\centering
\resizebox{9cm}{!}{\includegraphics[clip]{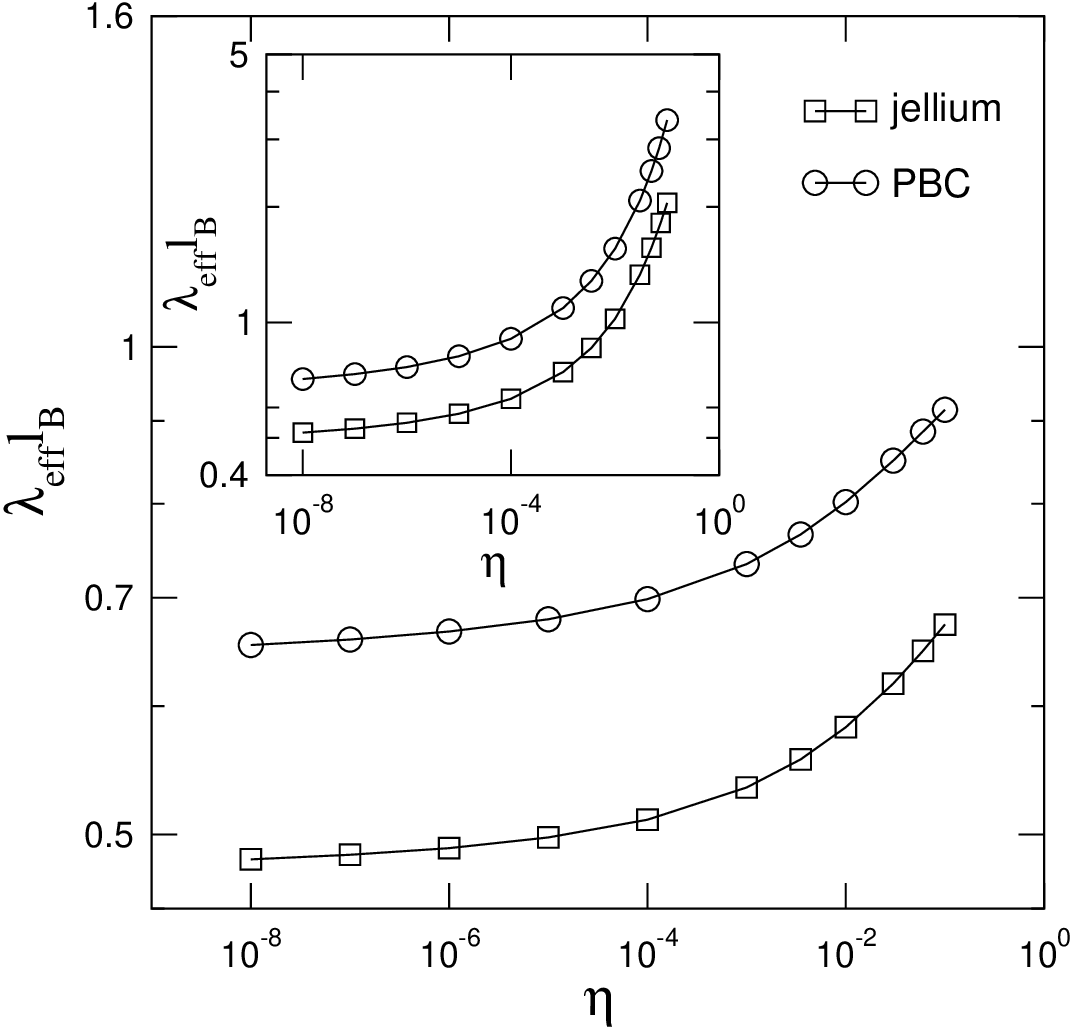}}
\caption{}
\label{fig:lambanosat}
\end{figure}

Fig. (\ref{fig:lambanosat}): Effective charge as a function of 
volume fraction within the PBC and the renormalized 
jellium  model, for $\lambda_{\bare}\ell_B=1$. The inset corresponds to the saturation
regime where $\lambda_{bare}\to\infty$ (rod-like colloids, no salt).\\

\begin{figure}[h]
\centering
\resizebox{9cm}{!}{\includegraphics[clip]{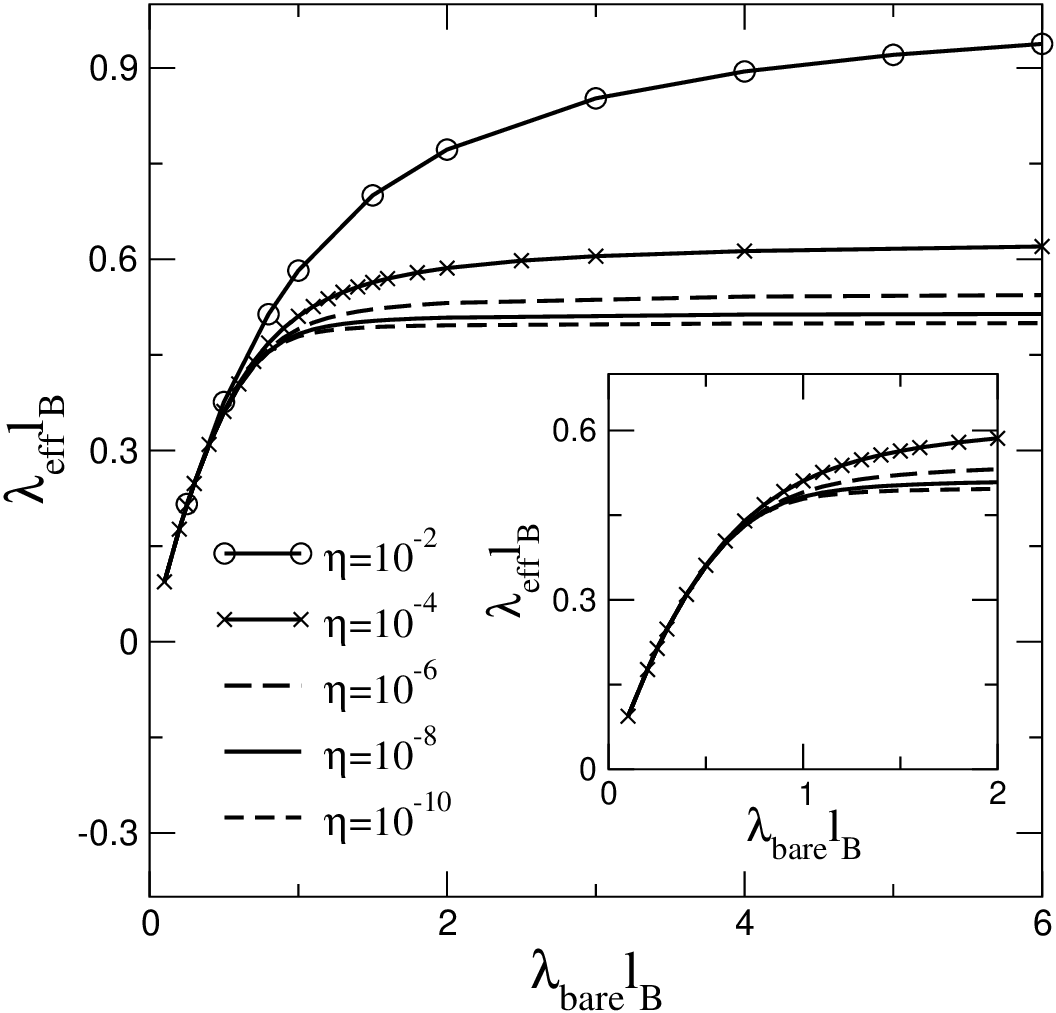}}
\caption{}
\label{fig:Manning}
\end{figure}

Fig. (\ref{fig:Manning}): The effective charge as a function of the bare charge for
different values of volume fractions (renormalized jellium model
for charged rods). The inset shows a magnification of the
main graph in the low charge
regime. The present scenario is exactly that of the Manning-Oosawa
counterion condensation occurring in the cell model.\\

\begin{figure}[h]
\centering
\resizebox{9cm}{!}{\includegraphics[clip]{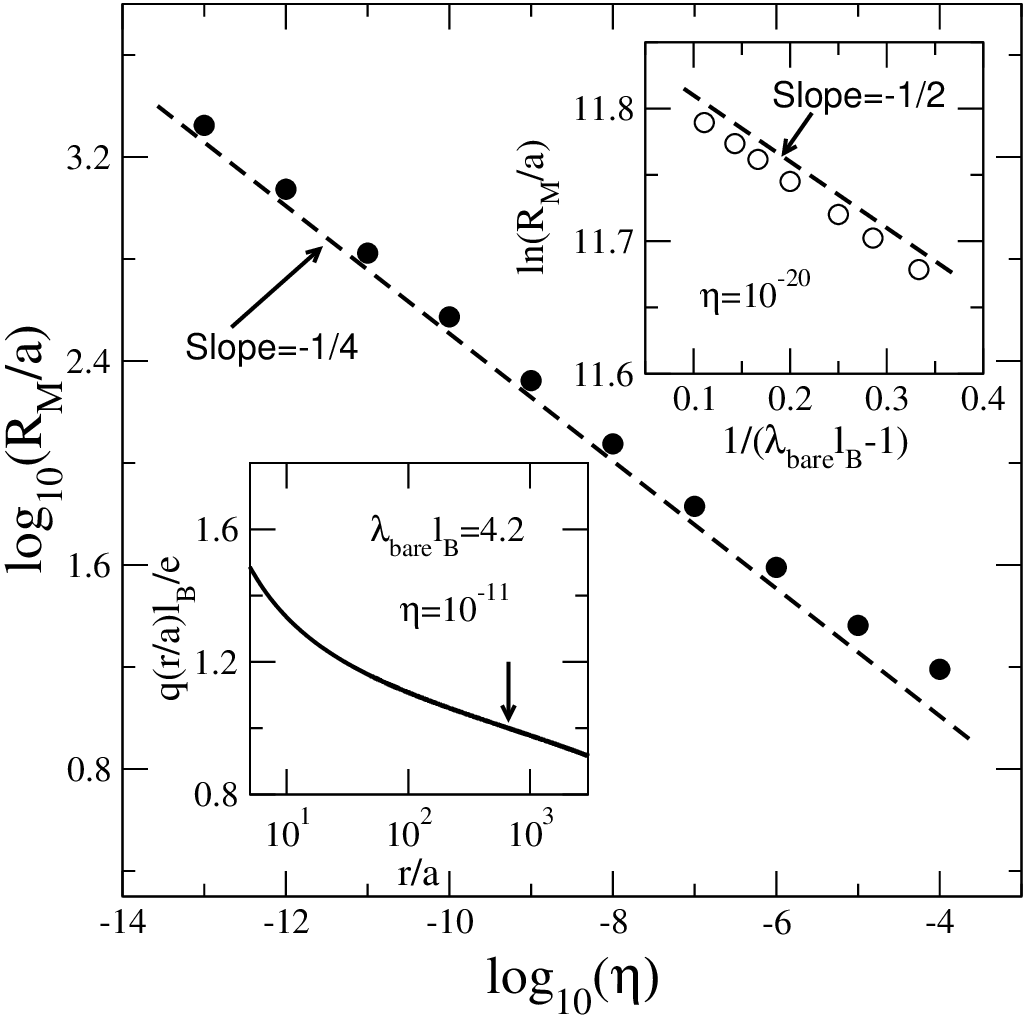}}
\caption{}
\label{fig:man1}
\end{figure}

Fig. (\ref{fig:man1}): Manning radius $R_M$ versus packing fraction for
$\lambda_\bare \ell_B = 4.2$. Extremely low densities have
been considered to see the predicted 
power law dependence $R_M\propto \eta^{-1/4}$, 
see Eq. (\ref{eq:mann}). The upper inset shows that the
bare charge dependence of $R_M$ also follows
the form given by Eq. (\ref{eq:mann}). The lower inset shows
$q(\tilde r)\ell_B/e$ as a
function of distance from the rod axis on a 
linear-log scale: as expected, the inflection point, 
indicated by the vertical arrow, coincides with the point where
$q(\tilde R_M)\ell_B/e=1$.\\

\begin{figure}[h]
\centering
\resizebox{9cm}{!}{\includegraphics[clip]{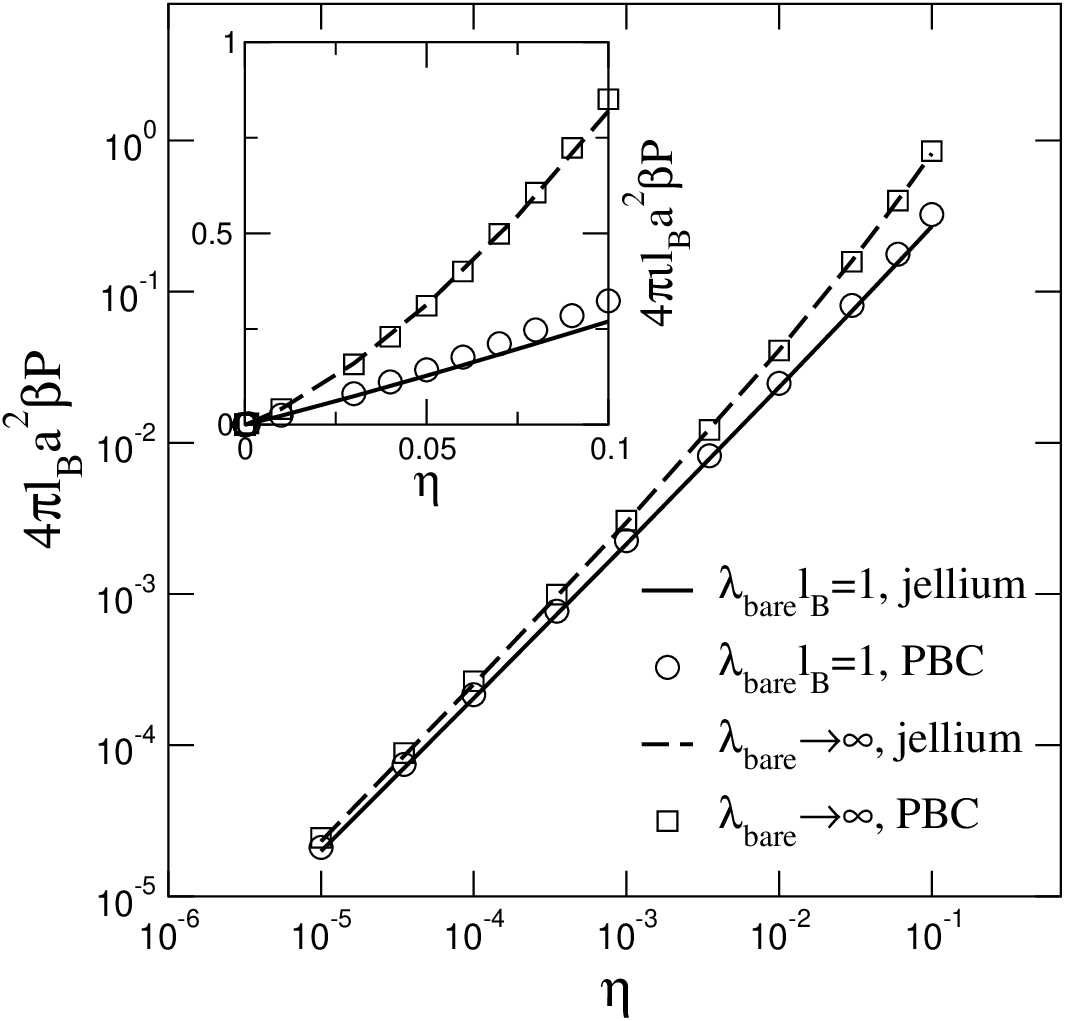}}
\caption{}
\label{press}
\end{figure}

Fig. (\ref{press}): Pressure as a function of volume fraction within the PBC and
the renormalized jellium model, for both a moderately charged, and a highly charged rods 
(saturation limit), without added salt. 
The inset shows the same data on a linear scale.\\

\begin{figure}[h]
\centering
\resizebox{9cm}{!}{\includegraphics[clip]{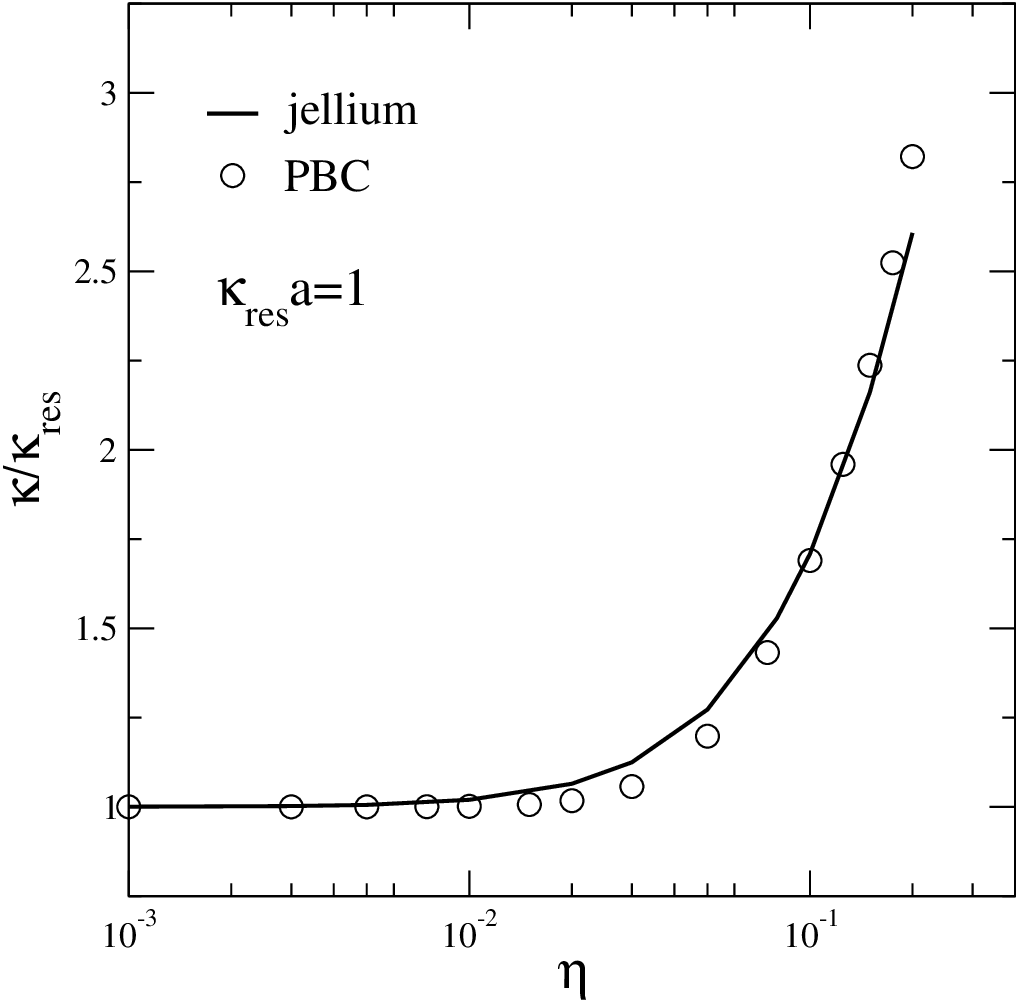}}
\caption{}
\label{fig:kapsphsalt}
\end{figure}

Fig. (\ref{fig:kapsphsalt}): Ratio between $\kappa $ and $\kappa_{\res}$ as a function of volume
fraction for $\kappa_{\res}a=1$ (saturation regime and spherical colloids).\\

\begin{figure}[h]
\centering
\resizebox{9cm}{!}{\includegraphics[clip]{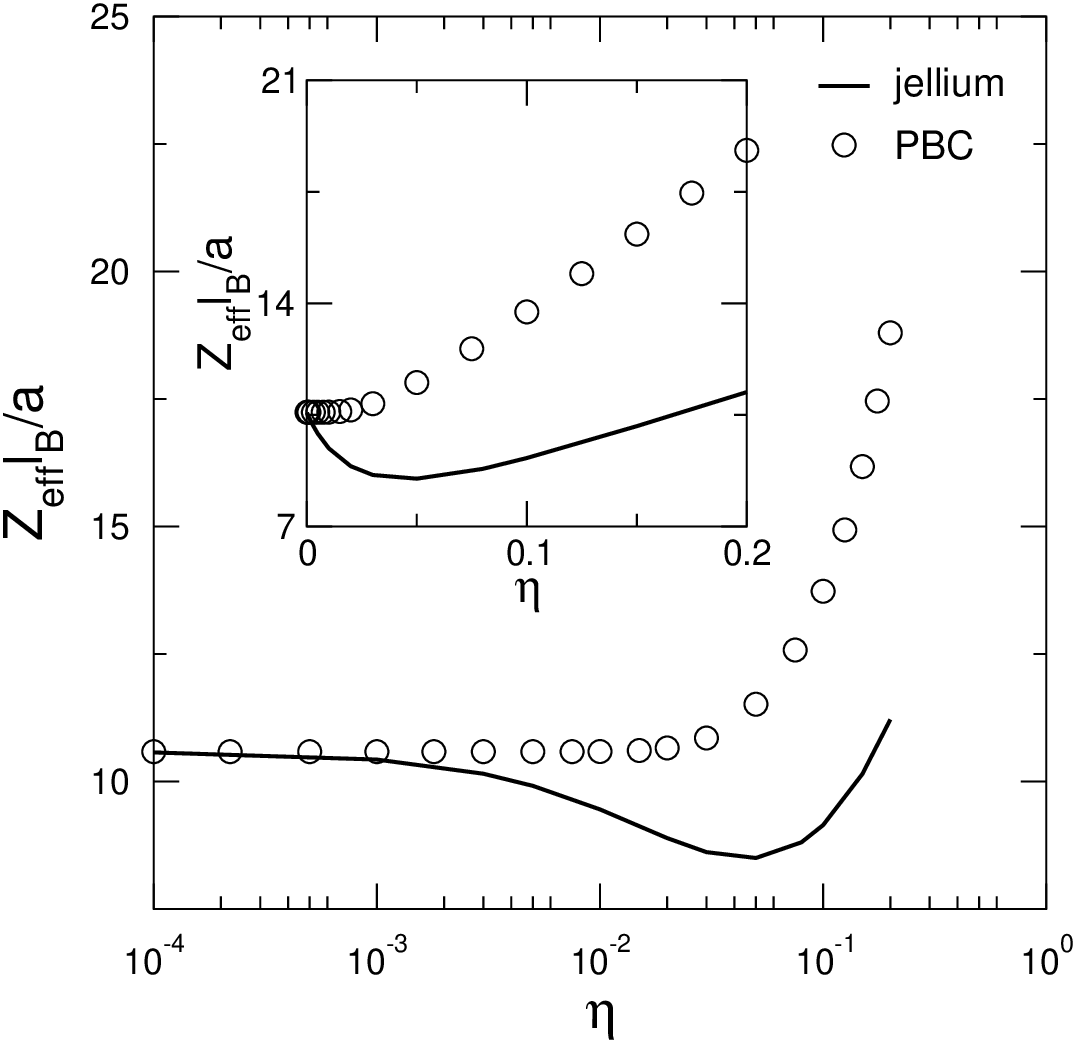}}
\caption{}
\label{fig:Zsphsalt}
\end{figure}

Fig. (\ref{fig:Zsphsalt}): The effective charge for spheres as a function of volume fraction
within the PBC and the renormalized jellium model for $\kappa_{\res}a=1$ in
the saturation regime $Z_\bare \to \infty$. The inset shows the same 
quantity on a linear scale.\\

\begin{figure}[h]
\centering
\resizebox{9cm}{!}{\includegraphics[clip]{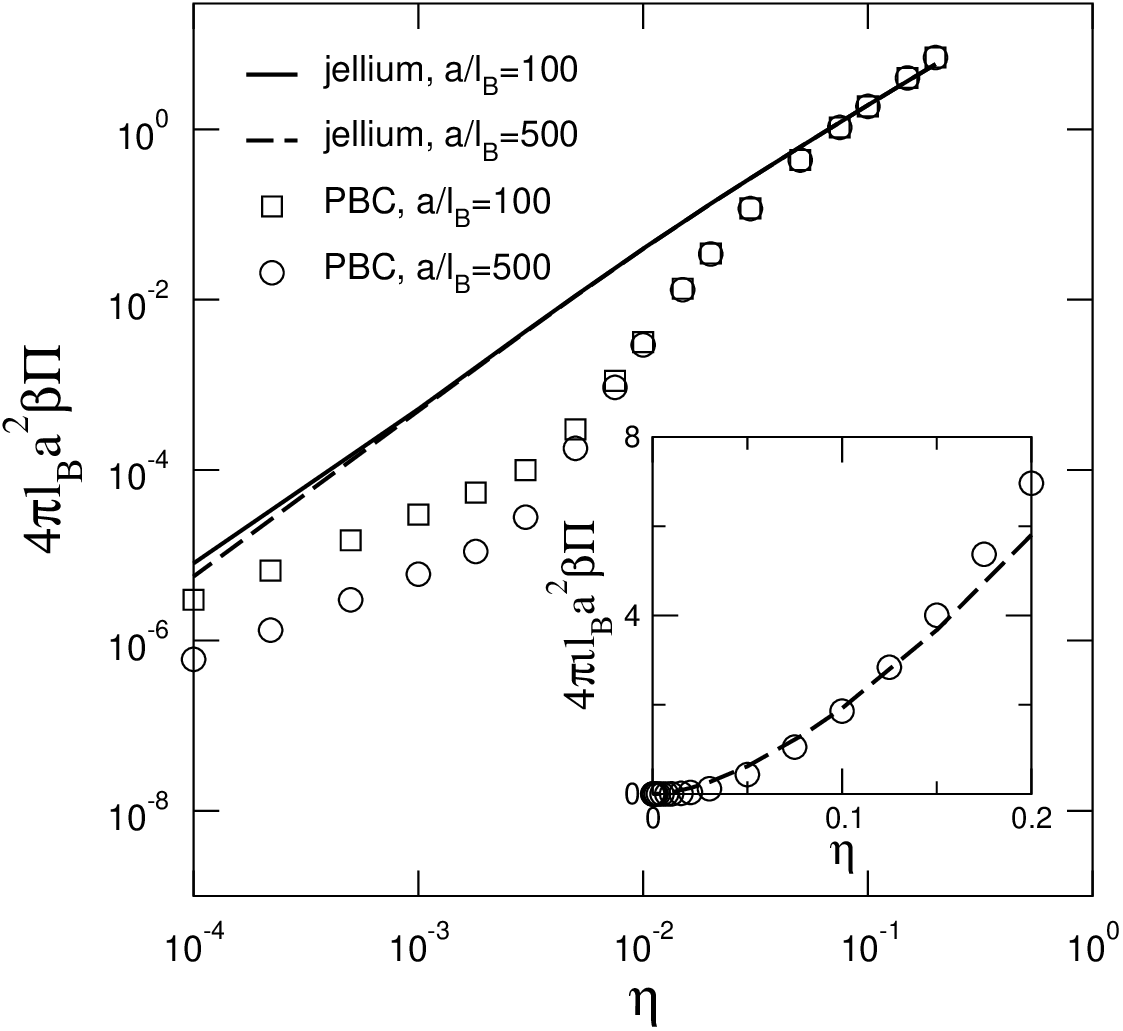}}
\caption{}
\label{fig:pressphsel}
\end{figure}

Fig. (\ref{fig:pressphsel}): The osmotic pressure as a function of volume fraction within the PBC
and the renormalized jellium model in the saturation regime for $\kappa_{\res}a=1$. The
inset  shows the same data on linear scale (spherical colloids).\\

\begin{figure}[h]
\centering
\resizebox{9cm}{!}{\includegraphics[clip]{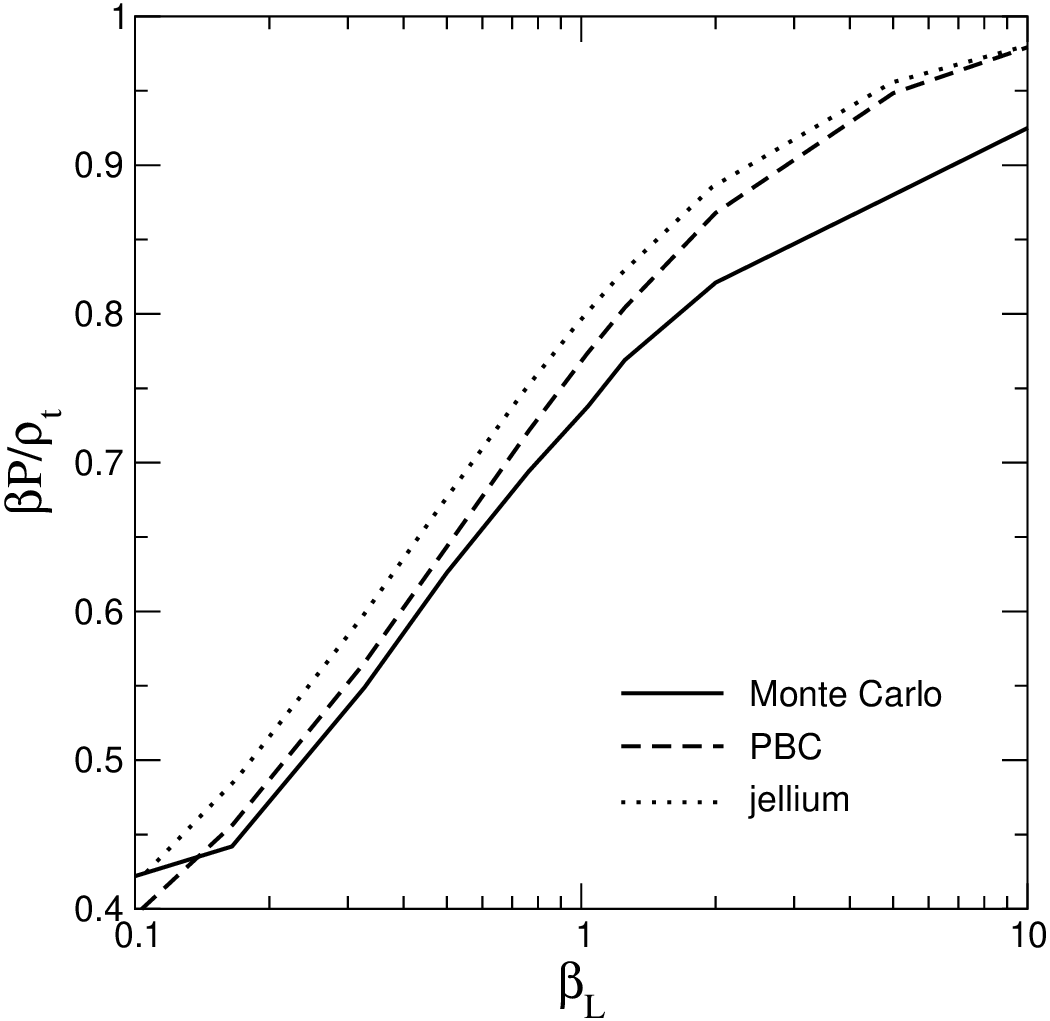}}
\caption{}
\label{fig:Lobaskin}
\end{figure}

Fig. (\ref{fig:Lobaskin}): Comparison of the PBC and the renormalized jellium  equations of state
with the one obtained in Ref. \cite{Lobaskin} from the Monte Carlo 
simulations. Here, the macro-ion volume fraction
is $\eta=8.4 \,10^{-3}$ while $Z_{\bare} \ell_B/a \simeq 21.45$.\\

\begin{figure}[h]
\centering
\resizebox{9cm}{!}{\includegraphics[clip]{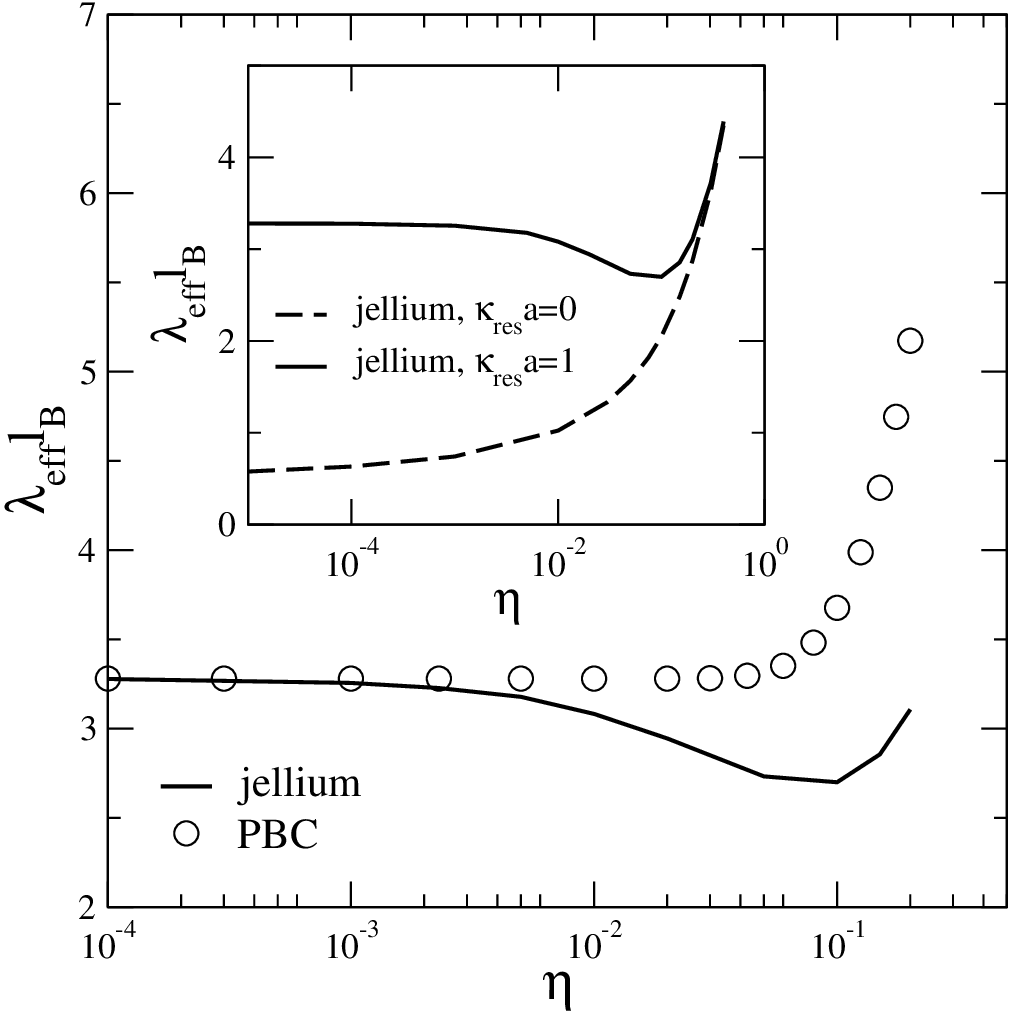}}
\caption{}
\label{fig:k0k1}
\end{figure}

Fig. (\ref{fig:k0k1}): The effective charge for highly charged cylindrical colloids
(saturation regime) as a function of volume
fraction within the PBC and the renormalized jellium model, for $\kappa_{\res}a=1$. 
The inset shows appearance of a minimum in the presence of salt.\\

\begin{figure}[h]
\centering
\resizebox{9cm}{!}{\includegraphics[clip]{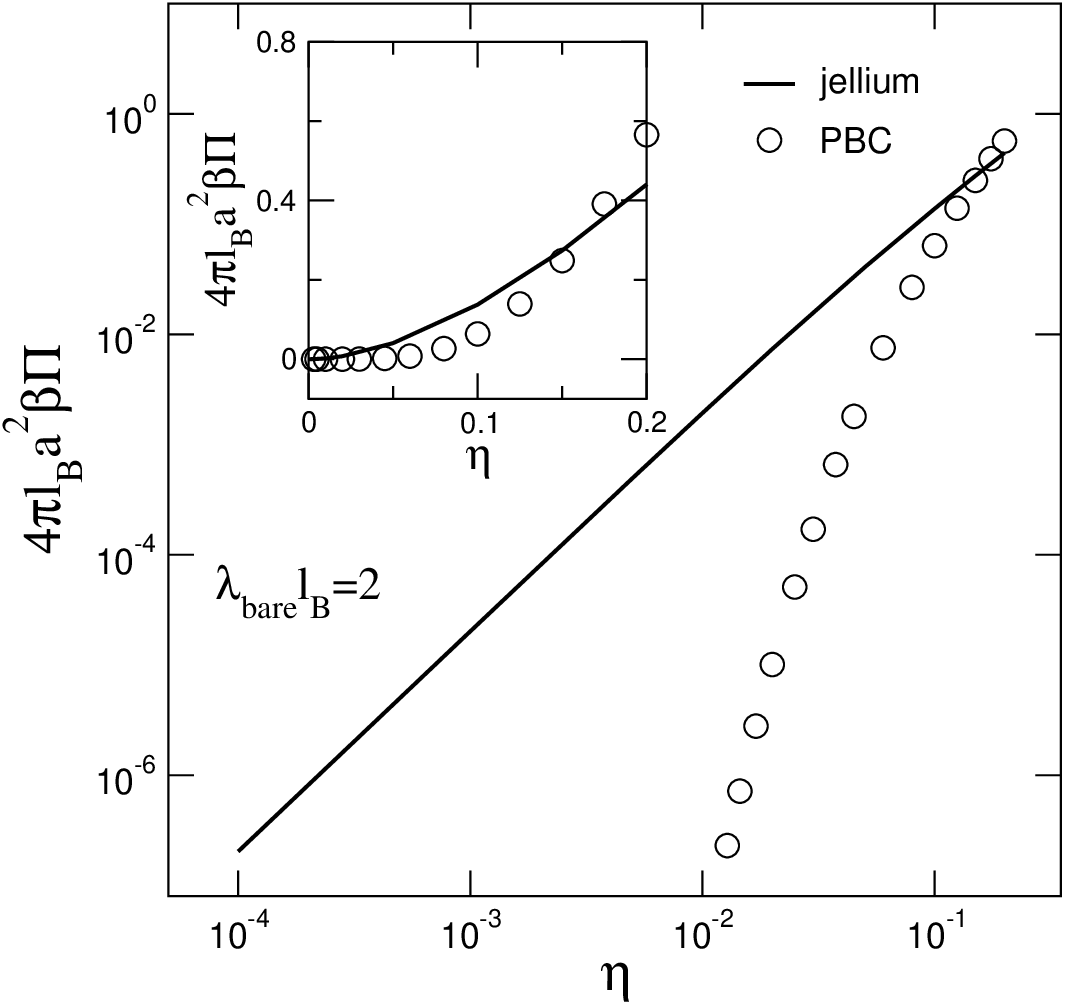}}
\caption{}
\label{presscylsel2}
\end{figure}

Fig. (\ref{presscylsel2}): Osmotic pressure as a function of volume fraction
within the PBC and the renormalized jellium model
for $\kappa_{\res}a=1$ and 
$\lambda_{bare}l_{B}=2$. The inset shows the same data on linear
scale (cylindrical colloids).\\


\newpage

\begin{thebibliography}{99}


\bibitem{Belloni} 
L. Belloni, 
J. Phys. Condens. Matter {\bf 12}, R549 (2000).

\bibitem{Alfrey}
T. Alfrey Jr, P.W. Berg  and H. Morawetz, 
J Polym. Sci. {\bf 7}, 543 (1951);      
R.M. Fuoss, A. Katchalsky and S. Lifson, 
P. Natl. Acad. Sci. USA {\bf 37}, 579 (1951).

\bibitem{Alexander}
S. Alexander, P.M. Chaikin, P. Grant, G.J. Morales, P. Pincus, and D. Hone, 
J.  Chem. Phys. {\bf 80},  5776  (1984).

\bibitem{Hansenrevue}  
J.-P. Hansen and H. L\"{o}wen, 
Annu. Rev. Phys. Chem. {\bf 51}, 209 (2000).

\bibitem{Deserno}
M. Deserno and C. Holm,  in {\em Proceedings of NATO Advanced Study Institute
  on Electrostatic Effects in Soft Matter and Biophysics}, edited by C. Holm,
  P. Kekicheff, and R. Podgornik (Kluwer, Drodrecht, 2001), p.\ 27.

\bibitem{Kuwabara}
S. Kuwabara,
J. Phys. Soc. Japan {\bf 14}, 527 (1959).

\bibitem{Ohshima}
H. Ohshima,
Coll. Surf. B {\bf 38}, 139 (2004).

\bibitem{LevinJPCM}
Y. Levin, E. Trizac and L. Bocquet,
J. Phys.: Condens. Matt. {\bf 15}, S3523 (2003).

\bibitem{LevinL} Y. Levin, M.C. Barbosa, M.N. Tamashiro   
Europhys. Lett.  {\bf 41}, 123 (1998); 
A. Diehl,  M. C. Barbosa, and Y. Levin, 
 Europhys. Lett.  {\bf 53}, 86 (2001)

\bibitem{Beresford}
B. Beresford-Smith, D.Y. Chan and D.J. Mitchell,
J. Colloid Interface Sci. {\bf 105}, 216 (1984);

\bibitem{Jellium}
E. Trizac and Y. Levin, 
Phys. Rev. E {\bf 69}, 031403 (2004).

\bibitem{Haro}
C. Haro-P\'{e}rez, M. Quesada-P\'{e}rez, J. Callejas-Fern\'{a}ndez, R. Sabate, J. Estelrich, 
R. Hidalgo-\'{A}lvarez,
Coll. Surfaces A {\bf 270}, 352 (2005).

\bibitem{HaroJFCM06}
C. Haro-P\'{e}rez, M. Quesada-P\'{e}rez, J. Callejas-Fern\'{a}ndez,
P. Schurtenberger, R. Hidalgo-\'{A}lvarez,
J. Phys.: Condens. Matter {\bf 18}, L363 (2006).

\bibitem{CastanedaCM}
R. Casta\~{n}eda-Priego, L. F. Rojas-Ochoa, V. Lobaskin, J. C. Mixteco-S\'{a}nchez,
Cond-mat/0608163.


\bibitem{Manning} 
G.S. Manning, 
J.~Chem.~Phys.~\textbf{51}, 924 (1969) ; \textbf{51}, 934 (1969);
F.~Oosawa, 
Polyelectrolytes, Dekker, New York (1971).

\bibitem{Manning2006}
E. Trizac and G. T\'ellez, 
Phys. Rev. Lett. {\bf 96}, 038302 (2006).

\bibitem{Levin}
Y. Levin, 
Rep. Prog. Phys. {\bf 65}, 1577 (2002).

\bibitem{Fushiki}
M. Fushiki, J. Chem. Phys. {\bf 97},  6700  (1992);
H. L\"owen, J.P. Hansen, and P.A. Madden, J. Chem. Phys. {\bf 98},  3275
(1993).

\bibitem{Graz}
  J. Dobnikar, Y.Chen, R. Rzehak and H.~H.~von~Gr\"unberg, 
  J. Chem. Phys. {\bf 119},4971 (2003);
  J. Dobnikar, D. Halo\v{z}an, M. Brumen, H.~H.~von~Gr\"unberg
  and R. Rzehak, 
  Comput. Phys. Commun. {\bf 159}, 73 (2004).
  
\bibitem{Belloni98}
L. Belloni, 
Colloid Surf. A {\bf 140}, 227 (1998).  

\bibitem{PRL2002}
E. Trizac, L. Bocquet and M. Aubouy,
Phys. Rev. Lett. {\bf 89}, 248301 (2002).

\bibitem{JCP2002}
 L. Bocquet, E. Trizac, and M. Aubouy, 
 J. Chem. Phys. {\bf 117}, 8138 (2002).

\bibitem{rque}
More precisely, it is the product $n_i^0 \exp(-\beta z_i e \varphi_\infty)$ 
which is physically relevant. One may always chose 
$\varphi_\infty=0$, modulo a proper redefinition of the pre-factors
$n_i^0$. This convention is convenient for salt free systems,
but has not been adopted in practice in the presence of added
salt. In addition, Eq. (\ref{eq:electroneutrality}) appears to be an effective
electro-neutrality condition, which does not coincide with the physical one.
This aspect is discussed at the beginning of section \ref{sec:salt}.

\bibitem{Langmuir2003}
E. Trizac, M. Aubouy, L. Bocquet, and H.H. von Gr\"unberg, 
Langmuir {\bf 19}, 4027 (2003).

\bibitem{rque2}
We emphasize that not all trial values of $\phi(\widetilde R)$ lead to 
a solution. For a given $\widetilde R$, 
there indeed exists a critical threshold $\phi^\sat(\widetilde R)$
beyond which no solution can be found. For small $\phi(\widetilde R)$
[i.e. $\phi(\widetilde R) \ll \phi^\sat(\widetilde R)$],
there is a linear relationship between $Z_\bare$ and $\phi(\widetilde R)$,
but when $\phi(\widetilde R)$ approaches $\phi^\sat(\widetilde R)$ from below,
the bare charge diverges. This is a consequence of the phenomenon
of effective charge saturation \cite{JCP2002,Tellezsat} that is ubiquitous
in mean-field treatments.

\bibitem{Tellezsat}
G. T\'ellez and E. Trizac, 
Phys. Rev. E {\bf 68}, 061401, 2003.

\bibitem{Gueron}
M. Gueron and G. Weisbuch,
Biopolymers {\bf 19}, 353 (1980).

\bibitem{OS}
B. O'Shaughnessy and Q. Yang,
Phys. Rev. Lett. {\bf 94}, 048302 (2005).

\bibitem{rque3}
This mean-field approach discards correlations between microions,
that become prevalent at large electrostatic couplings 
(see \cite{Grosberg}). In a solvent like water 
at room temperature, PB theory nevertheless provides a good description 
of monovalent ion systems, see e.g. \cite{DesernoMay} or
\cite{Levin}. 

\bibitem{Grosberg}
A.Y. Grosberg, T.T. Nguyen and B.I. Shklovskii,
Rev. Mod. Phys. {\bf 74}, 329 (2002).

\bibitem{DesernoMay}
M. Deserno, C. Holm and S. May,
Macromolecules {\bf 33}, 199 (2000).

\bibitem{Lobaskin}
V. Lobaskin and K. Qamhieh,
J. Phys. Chem. B {\bf 107}, 8022 (2003).



\end{thebibliography}
\end{document}